# Molecular dynamics simulation of nanocolloidal amorphous silica particles
# Part II

S. Jenkins and  S.R. Kirk,
*Dept. of Technology, Mathematics & CS,*
*University West, P.O. Box 957, Trollhättan, SE 461 29, Sweden.*

M. Persson and J. Carlen,
*R&D Pulp and Paper, Eka Chemicals (Akzo Nobel) AB, Rollsbo, Sweden.*

Z. Abbas
*Department of Chemistry, Göteborg University, SE-412 96 Gothenburg Sweden.*

**Abstract:** Explicit molecular dynamics simulations were applied to a pair of amorphous silica nanoparticles of diameter 3.2 nm immersed in a background electrolyte. Mean forces acting between the pair of silica nanoparticles were extracted at four different background electrolyte concentrations. Dependence of the inter-particle potential of mean force on the separation and the silicon to sodium ratio, as well as on the background electrolyte concentration, are demonstrated. The pH was indirectly accounted for via the ratio of silicon to sodium used in the simulations. The nature of the interaction of the counter-ions with charged silica surface sites (deprotonated silanols) was also investigated. The effect of the sodium double layer on the water ordering was investigated for three Si:Na$^+$ ratios. The number of water molecules trapped inside the nanoparticles was investigated as the Si:Na$^+$ ratio was varied. Differences in this number between the two nanoparticles in the simulations are attributed to differences in the calculated electric dipole moment. The implications of the form of the potentials for aggregation are also discussed.

**Introduction**

Silica in colloidal form [1] has been used in industrial applications and studied for decades, both in terms of production techniques [2,3] and methods for functionalization. In addition to the more traditional range of applications (e.g. in the food, paint, coatings and paper industries), new ranges of applications in, e.g. the biomedical industries [4,5] have been developed. In addition, grafted polymeric species on nanocolloidal silica enhance dispersant properties [6], silica nanoparticles can be used as hosts and carriers for other smaller, possibly toxic, particulates [7], and silica nanocolloids also form the basis for the production of zeolites [8,9], which have a wide range of economically significant applications. In these applications, the ability to control the stability of silica nanocolloids in (usually) aequous solutions, both in terms of the chemical stability of the particles themselves, and also colloidal stability (through the interactions between the particles themselves, and possibly any other chemical species present), is vital in order to optimise their effectiveness in the desired application.

The Derjaguin-Landau-Verwey-Overbeek (DLVO) theory of interparticle interactions [10-12], containing terms consisting of a balance of attractive van der Waals forces and repulsive electrical double-layer forces, has been widely and successfully used to predict stability in a wide range of colloidal systems. This theory, however, does not account for specific ion effects, and, more importantly, provides incorrect predictions of both chemical stability and colloidal aggregation of silica particles up to sub-micrometer sizes in environments with high levels of background salt and/or low pH [13,14]. Aggregation processes in silica colloids are strongly influenced by solution pH [1], and the range of potential applications spans a wide range of solution pH values (and background electrolyte concentrations), so it is essential that the effects of pH on these systems be thoroughly understood.

Theoretical models for property prediction in nanocolloidal systems have been developed, covering a wide range of length and time scales at different levels of theory [15,16]. Of particular interest in this paper is the use of molecular dynamics simulations to probe the detailed interactions of nanocolloidal particles with each other, solvent molecules and counter-ions in solution. It is already well known that pH is a significant factor influencing both the growth [1,17] and colloidal stability [1,18] of colloidal nanosilica and, indeed, colloids in general. The pH of a system has, until recent years, been difficult to account for directly in MD simulations. Several approaches to the problem have recently been made for implicit solvent [19] and explicit solvent [20-22] approaches, culminating in the latter case in so-called 'acidostat' methods [23,24]. In our case, we have investigated silica nanoparticles produced by cation exchange of a sodium silicate solution, and the pH has been indirectly accounted for via the ratio of silicon to sodium present in the simulations [1]. The number of deprotonated silanol sites on the surface of the modelled silica nanoparticles has therefore been set equal in all cases to the number of associated sodium cations (related via the fixed Si:Na$^+$ ratio to the number of silicon atoms in the simulation), to ensure overall electroneutrality. We have performed our simulations for a number of industrially-relevant ranges of the silicon to sodium (Si:Na$^+$) ratio, which we use to indirectly account for varying pH (see Methods and Results sections for more details and the companion paper [25] to this current work, where the details of the modelling of amorphous silica particle structure and allocation of surface charge on silica test particles are provided).

## II. METHODS

Details of the methods used to create the amorphous silica nanoparticle structure and the algorithm used for the allocation of surface charge on the particles can be found in the companion paper to this current work [25]. Three different Si:Na$^+$ ratios were chosen in this work, corresponding to realistic values used in the industrial process for the silica nanoparticles, specifically 20:1, 10:1 and 5:1.

**Molecular dynamics**

Molecular dynamics calculations were carried out in the NVT ensemble using the GROMACS [26-28] code (version 3.3.1), using OPLS-AA [29] force fields, with additional parameters for silica from literature [30]. The water model used was a flexible variant of the TIP4P 4-site model [31,32]. During actual MD runs, the simulation temperature of all species was set to 300K, particle velocities being chosen from a Maxwell distribution, using a Berendsen [33] thermostat with a coupling time constant of 0.1 ps. Neighbour list, Coulomb and van der Waals cutoffs were all set to 1 nm. Centre of mass motion was removed after every timestep. Long-range electrostatic forces were treated using a particle-mesh Ewald [34] treatment, with long-range dispersion corrections applied. The FFT grid spacing was in all cases around 0.118 nm, varying slightly with simulation box size, and cubic (order 4) spline interpolation was used on the FFT grid. In all cases, a 3-stage MD protocol was used - energy minimization, followed by 100 ps (0.002 ps timestep) of position-restrained MD, followed by the 'production' run (performed with 0.001 ps timestep). In cases where a non-zero background salt concentration was modelled, an appropriate number of water molecules in the simulation box were randomly substituted with equal numbers of $Na^+$ and $Cl^-$ ions to reach the desired concentration.

**Single-particle MD runs**

The test nanoparticles were centered in cubic simulation boxes with periodic boundary conditions; the boxes were then filled with water, the box size being chosen so that there was at least 1 nm of water between any side of the box and the nanoparticle, in order to avoid potential spurious water structuring effects caused by the periodicity of the simulation box [35]. Sodium ions were substituted in randomly for water molecules until the contents of the simulation box was overall electrically neutral. The production run in these cases consisted of a 500 ps unconstrained MD run, all other parameters being as previously stated.

**Double-particle MD runs - PMF calculations**

The fully equilibrated test nanoparticles from the single-particle MD runs were used as building blocks to investigate interparticle interactions. Pairs of nanoparticles (along with their accompanying cloud of neutralizing $Na^+$ ions) were placed relative to each other in such a way that the distance between their centres of mass (COM distance) took a number of specified values. The resulting nanoparticle pairs were then centered in large simulation boxes with periodic boundary conditions applied. The dimensions of these large simulation boxes were chosen such that the any part of one nanoparticle was closer to all parts of the other member of the pair than to any periodic copies of either particle. The remaining space in the simulation boxes was filled with TIP4P water. In these runs the production phase consisted of a 500 ps (0.001 ps time-step) potential of mean force (PMF) calculation, using the constraints method built into the pull code within GROMACS. The COM distance between the nanoparticle pair at the end of the position-restrained MD run was used as a constraint distance. An additional constraint force was applied by the GROMACS code during the PMF production run to maintain this original COM distance, and the value of the constraint force needed to maintain this distance was monitored and recorded at every time-step of the production run. After the end of the production run, the 500,000 values of these constraint forces were averaged, representing the average interparticle attractive or repulsive force caused by interactions between the particles [36]. Variants of these basic PMF runs were generated and carried out for various different background concentrations of NaCl in the solvent (the additional ions having been present from the beginning of the protocol), in the form of additional $Na^+$ and $Cl^-$ ions randomly substituted for solvent (water) molecules. Numbers of additional ions were chosen to replicate molecular background salt concentrations of 0.00M, 0.01M, 0.1M and 1.0M, encompassing the range of background salt levels used in industrial production of these nanoparticles, and found in their common applications.

**Results and discussion**

The amorphous silica nanoparticles consist mainly of tetrahedrally coordinated silicon atoms, including silanol bridges, surface hydroxyl groups and deprotonated oxygen surface ($O^-$) sites. For one particle size, the number of surface sites is determined by the $Si:Na^+$ ratio; several ratios were chosen (5:1, 10:1 and 20:1) in this work. For these three $Si:Na^+$ ratios, the number of charged oxygen surface sites (also equal to the number of sodium counter ions) was 21, 42 and 84 per nanoparticle respectively.

Several features of the surface topology observed in this work were found to be rather similar to those observed in a companion study (part I of this work [25]) conducted on a 4.4 nm particle with a 20:1 $Si:Na^+$ ratio, and so have not been repeated here, including snapshots of the local arrangement of the water molecules around a deprotonated oxygen surface site, the various hydrogen bonding length (hydrogen-acceptor) distributions and form (but not the actual ratios) of radial distribution functions (*rdf*) for water around the sodium ions. As observed in the previous work, there is a lack of strong chemical bonding of the sodium (or surface adsorption) to the deprotonated silanol surface site ($O^-$), which has important consequences for the aggregation properties of the silica nanoparticles.

In Figures 1(a-c) the effect of Si:$Na^+$ ratio on the water ordering is demonstrated. These figures show axial-radial density plots of the distribution of water molecules around the pair of nanoparticles in the presence of only the $Na^+$ counter-ions, for the entire MD trajectory; the axis corresponds to the line joining the centres of mass of the two particles. It can be seen that the effect of decreasing the Si:$Na^+$ ratio, that is to increase the number of sodium counter-ions, is to disrupt the water ordering around the nanoparticle. Notice the faint layering effect within approximately 0.3nm above the nanoparticle surface, that can be seen in Figure 1(b). This effect becomes more noticeable in Figure 1(c) but is not readily apparent in Figure 1(a). Figure 2(a-c) shows the corresponding plots for the sodium counter-ions around the pair of nanoparticles, which can be explained in terms of the changes in the water ordering with Si:$Na^+$ ratio, since additional quantities of $Na^+$ ions (with their accompanying hydration shells) will obstruct other water molecules forming hydrogen bonds with the surface $O^-$ sites. However, the additional $Na^+$ ions do not prevent water molecules from entering the surface of the nanoparticle, as can be seen from Table 1, where the number of water molecules entering each of the pair of nanoparticles and an isolated nanoparticle are presented for the three Si:$Na^+$ ratios of this study. There is a steady trend of increasing numbers of water molecules entering the double and single nanoparticles with decreasing Si:$Na^+$ ratio. This could be explained by the necessity that a sodium ion will lose some of its hydration shell as it approaches the nanoparticle surface, and then electrostatically interacts with the surface. As a consequence, some of the hydration shell molecules will become trapped inside the nanoparticle. From Table 1 it can be seen that there is a drop in the number of water molecules entering the nanoparticle at the maximum background $Na^+$ concentration (1.00M); this is likely due to the large numbers of background $Na^+$ ions interfering with those in the vicinity of the nanoparticles surface.

In order to quantify the spatial effects of the various background concentrations of $Na^+$ ions on the surface and beyond, we have calculated the radial distribution functions (*rdf*) for water around the sodium ions, in a similar manner as performed in Part I of this work [25], for all background salt concentrations considered. These showed the interesting feature that the *rdf* after the first peak is close to zero, which is more similar to the typical *rdf* of a solid rather than a liquid, showing as it does some ordering or layering effects in the water around the counter-ions closest to the nanoparticle surface.

In Table 3 a summary of the *rdf* data is presented for the three values of the Si:$Na^+$ ratio for the silica nanoparticles considered in this study, listing both the pairwise interaction of same-sized nanoparticles and the isolated nanoparticle for comparison. Examining the first set of entries ($Na^+$ and $H_2O$) in Table 3, it can be seen that there is a monotonic increase of the height ratio (second peak to first, used to quantify the extent to which the second species is 'piling up' around the first species used in the analysis) with increasing background sodium concentration. This holds true for both the double particles and isolated particles. Closer inspection shows that the *rdf* ratio entry for the double particle corresponding to Si:$Na^+$ ratio 5:1 is almost identical to that of the isolated particle. As the Si:$Na^+$ ratio increases, the differences between the double and isolated nanoparticles increase for a given Si:$Na^+$ ratio, especially at 0.00M background $Na^+$ ion concentration. The increase in the *rdf* ratio with increasing background sodium concentration could be explained in terms of the sodium ions binding the water and preventing it from entering the nanoparticle. In addition, the fact that the *rdf* ratio for the 5:1 Si:$Na^+$ ratio and 0.00M are very similar for the isolated and double particle cases may explain why the nanoparticle lets in so many water molecules; the sodium double layer for the pair of particles is no more effective than for the isolated nanoparticle at preventing water entering the surface of the nanoparticle.

In general, we observed that the multiple simulation runs started with initial COM distance values chosen with a constant increment (0.5 nm) yielded constraint distances for the PMF phase of the simulation runs (i.e. after the preliminary temperature equilibration position-restraints phase) which did not have a constant spacing, as can be seen from Figures 3-5, under the influence of the oscillatory interparticle forces. In Figure 6(a-c) we show the calculated potential of mean forces (PMF) for three sets of Si:$Na^+$ ratio; 20:1, 10:1 and 5:1 used in this study, by integration of the mean interparticle forces. We removed the constant of integration by estimating the potential at infinite range (i.e. zero) to be the value at the largest separation used in our calculations. A decrease in Si:$Na^+$ ratio makes the nanoparticles behave more like simple spherical particles with smooth hard surfaces, and separated by small spherical solvent molecules with respect to their solvation forces. The silica nanoparticles, however, are neither perfectly spherical, nor smooth, nor hard, so it is not surprising that their inter-particle forces do not show very smooth oscillations. This can be observed from the form of the solvation forces for a Si:$Na^+$ ratio of 5:1, which obey a decaying oscillatory function with particle separation, including both the attractive depletion and repulsive structural energy barrier [37]. This trend can be seen progressively increasing from the largest Si:$Na^+$ ratio through to the smallest, see Figures 3-5. A general trend in the *average* forces can be seen in Table 2, e.g. that the addition of 0.10M $Na^+$ is associated with overall attractive forces which increase monotonically as the Si:$Na^+$ ratio decreases, i.e. as the amount of $Na^+$ increases. The maximum repulsive forces (at the 0.10M $Na^+$ concentration) decrease monotonically as the Si:$Na^+$ ratio decreases, whilst the maximum attractive force is very similar for the 20:1 to 10:1 Si:$Na^+$ ratios and increases almost by a factor of six for the 5:1 Si:$Na^+$ ratio. In particular, notice that in Figure 5(c) for a Si:$Na^+$ ratio of 5:1 and a $Na^+$ concentration of 0.10M that the form of the oscillatory interparticle forces have a form suggesting that the solution-nanoparticle surface interaction is more attractive and hydrophobic [37] (at separations of greater than 3.8 nm) than the interaction in Figure 5(b) where there is a simple decaying oscillatory form. In Figure 4(d) where the Si:$Na^+$ ratio is 10:1 the surface is repulsive and hydrophilic. Examination of the corresponding interparticle potential for Figure 5(c) in Figure 6(c) for 0.10M shows that the solution-nanoparticle surface interaction is indeed attractive, more so than the potentials for the other three background sodium concentrations (in Figure 6c). Conversely, for a Si:$Na^+$ ratio of 10:1 and 1.00M background $Na^+$ concentration (shown in Figure 4(d)) the solution-nanoparticle surface interaction is repulsive, as can be seen from the form of the potential in Figure 6(b). Examples of interactions which show force curves that oscillate about the separation axis are shown in Figures 5(b) and 5(d) where the corresponding potentials show attraction and repulsion. The potentials become more attractive as the Si:$Na^+$ ratio decreases. The number of water molecules trapped inside the nanoparticles (shown in Table 1) for both isolated and double particle cases is correlated with the magnitude of the corresponding electric dipole moments (from Table 4), i.e. the larger the electric dipole moment, the greater number of water molecules are trapped inside the nanoparticle.

**Conclusions**

This study, along with the accompanying part I [25], has successfully modelled the interactions between, and near-surface structure of, amorphous silica nanoparticles in aqueous solutions with varying amounts of background counter-ions and Si:$Na^+$ ratios, including those of relevance to the industrial setting using realistic molecular dynamics force fields, evaluating the interparticle potentials using a PMF formalism. This work also shows that water molecules can penetrate into amorphous silica nanocolloid particles. We found, in conjunction with the accompanying part I [25], that 'hairiness' of the silica surface (modelled at atomic resolution with a realistic silica surface structure) has an effect on the interparticle potential of mean force, in agreement with other work in the literature [38,39], describing a thin 'hairy' layer on the surfaces of such systems. The consequences for water ordering have also been investigated, in conjunction with the presence of surface-bound and free background counter-ions, and we have quantified the factors influencing the effectiveness of the sodium counter-ion 'double layer' in preventing or enabling water molecules entering the silica nanoparticles. In addition, the number of water molecules trapped inside the nanoparticles was shown to be correlated with the values of the electric dipole moments.

Reactive MD potentials [40] and QM/MM [41,42] treatments may provide a future avenue for investigations of the longer time-scale interactions between, for instance, 'crashed' particles as condensation reactions between silanol groups create bonds as the particles 'fuse' together. At longer length and time-scales, the interparticle potentials derived in this work will inform ongoing coarse-grained MD investigations of flocculation and gelation in silica nanocolloid systems. In particular, the aggregation kinetics and morphologies produced by the sets of potentials generated by varying the available process parameters will be investigated, using, among other techniques, a fractal analysis (as demonstrated by Videcoq [43]) to characterise the aggregates produced by the different shaped potentials. From this it will be possible to select those potentials which lead to either fractal-like or spherical shaped aggregates; the ability to do so is of great interest for both industrial colloidal silica applications as well as colloid science. The eventual long term goal of this work is the creation of a microscopic theory of gelation, in partnership with novel application of a mathematical network formalism, utilizing the universal character of colloids [44].

**Acknowledgments**

The Knowledge Foundation are gratefully acknowledged for the support of SJ and SRK, grant number 2004/0284. This work was made possible by the facilities of the Shared Hierarchical Academic Research Computing Network (SHARCNET:www.sharcnet.ca), through the kind auspices of our sponsor Dr. P.W. Ayers, Department of Chemistry, McMaster University, Ontario, Canada.

**Table 1.** The mean of the time-averaged number of water molecules trapped inside nanoparticles.

| | Background sodium ion (Na$^+$) molarity | | | |
|---|---|---|---|---|
| | **0.00** | **0.01** | **0.10** | **1.00** |
| **Si:Na$^+$ ratio=20:1** | | | | |
| *snp 1* | 21.54 | 21.05 | 20.70 | 19.11 |
| *snp 2* | 22.61 | 22.59 | 22.79 | 20.74 |
| *isolated nanoparticle* | 19.70 | 19.20 | 18.30 | 18.30 |
| **Si:Na$^+$ ratio=10:1** | | | | |
| *snp 1* | 21.53 | 21.55 | 21.37 | 19.93 |
| *snp 2* | 21.47 | 21.58 | 21.11 | 19.71 |
| *isolated nanoparticle* | 22.60 | 22.00 | 22.70 | 18.60 |
| **Si:Na$^+$ ratio=5:1** | | | | |
| *snp 1* | 32.82 | 32.66 | 32.88 | 31.11 |
| *snp 2* | 33.62 | 33.33 | 32.68 | 31.05 |
| *isolated nanoparticle* | 33.60 | 32.70 | 32.10 | 26.50 |

**Table 2.** Variation of the inter-particle force for an entire MD trajectory with separation, background sodium concentration and Si:Na$^+$ ratio. (see Figures 3-5 for plots of the inter-particle forces vs. separation). The first figure under each concentration heading is the average force and the maximum attractive and repulsive values of the inter-particle forces are shown italicized, and the units used are kJ mol$^{-1}$ nm$^{-1}$.

| | Background sodium ion concentration(molarity) | | | |
|---|---|---|---|---|
| Si:Na$^+$ ratio | **0.00** | **0.01** | **0.10** | **1.00** |
| **20:1** | 7.945(*-76.19, 72.94*) | 10.53(*-48.42, 80.52*) | -1.865(*-70.17, 125.6*) | 19.00(*-46.81, 162.6*) |
| **10:1** | 36.08(*-43.00, 276.6*) | 13.65(*-54.63, 260.6*) | -60.13(*-70.56, 108.1*) | 28.41(*-111.3, 303.8*) |
| **5:1** | 1.451(*-195.4, 271.8*) | -14.16(*-239.9, 247.1*) | -73.87(*-408.7, 70.15*) | 7.325(*-319.3, 390.1*) |

**Table 3.** Summary of the radial distribution function peak height ratios for the silica nanoparticle(s) with diameter 3.2 nm. Each entry in the table is the ratio of the heights of the second to first peaks of the corresponding radial distribution function for each background sodium concentration given by the column heading (see also Figure 3). 'snp 1' and 'snp 2' refer to individual results for the two silica nanoparticles when in a pair. The two nanoparticles are distinguishable, see Tables 3 and 4.

| | Background sodium ion ($Na^+$) molarity | | | |
|---|---|---|---|---|
| **RDF species** | **0.00** | **0.01** | **0.10** | **1.00** |
| **Si:$Na^+$ ratio=20:1** | | | | |
| **$Na^+$ and $H_2O$** | | | | |
| *snp 1 + snp 2* | 0.357 | 0.380 | 0.454 | 0.564 |
| *isolated nanoparticle* | 0.472 | 0.452 | 0.496 | 0.567 |
| **Si:$Na^+$ ratio=10:1** | | | | |
| **$Na^+$ and $H_2O$** | | | | |
| *snp 1 + snp 2* | 0.387 | 0.397 | 0.475 | 0.564 |
| *isolated nanoparticle* | 0.370 | 0.414 | 0.471 | 0.558 |
| **Si:$Na^+$ ratio=5:1** | | | | |
| | **0.00** | **0.01** | **0.10** | **1.00** |
| **$Na^+$ and $H_2O$** | | | | |
| *snp 1 + snp 2* | 0.401 | 0.403 | 0.459 | 0.556 |
| *isolated nanoparticle* | 0.402 | 0.416 | 0.444 | 0.560 |

**Table 4.** Variation of the average nanoparticle electric dipole moment with Si:$Na^+$ ratio; see the caption for Table 3 for explanation of terms.

| | Average nanoparticle dipole moment/Debye |
|---|---|
| **Si:$Na^+$ ratio=20:1** | |
| *snp 1* | $1.0 \times 10^4$ |
| *snp 2* | $1.3 \times 10^4$ |
| *isolated nanoparticle* | $8.6 \times 10^3$ |
| **Si:$Na^+$ ratio=10:1** | |
| *snp 1* | $2.2 \times 10^4$ |
| *snp 2* | $2.7 \times 10^4$ |
| *isolated nanoparticle* | $1.7 \times 10^4$ |
| **Si:$Na^+$ ratio=5:1** | |
| *snp 1* | $4.5 \times 10^4$ |
| *snp 2* | $5.5 \times 10^4$ |
| *isolated nanoparticle* | $3.5 \times 10^4$ |

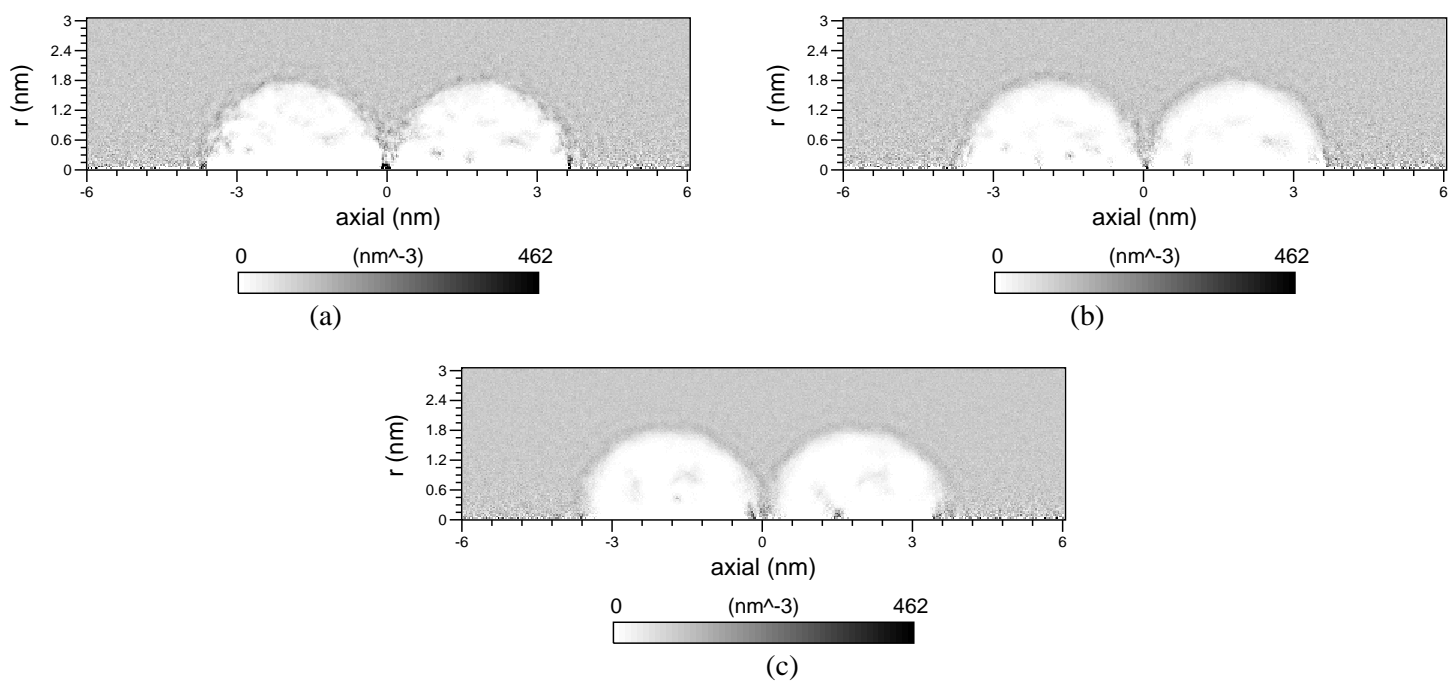

**Figure 1.** The axial-radial density plot of the distribution of water molecules around the pair of nanoparticles for the entire MD trajectory, where the sub-figure (a) corresponds to the result for Si:Na$^+$ ratio of 5:1, (b) 10:1 and (c) 20:1. The axis corresponds to the line joining the centres of mass of the two particles. The same density scale is used for all three plots for ease of comparison.

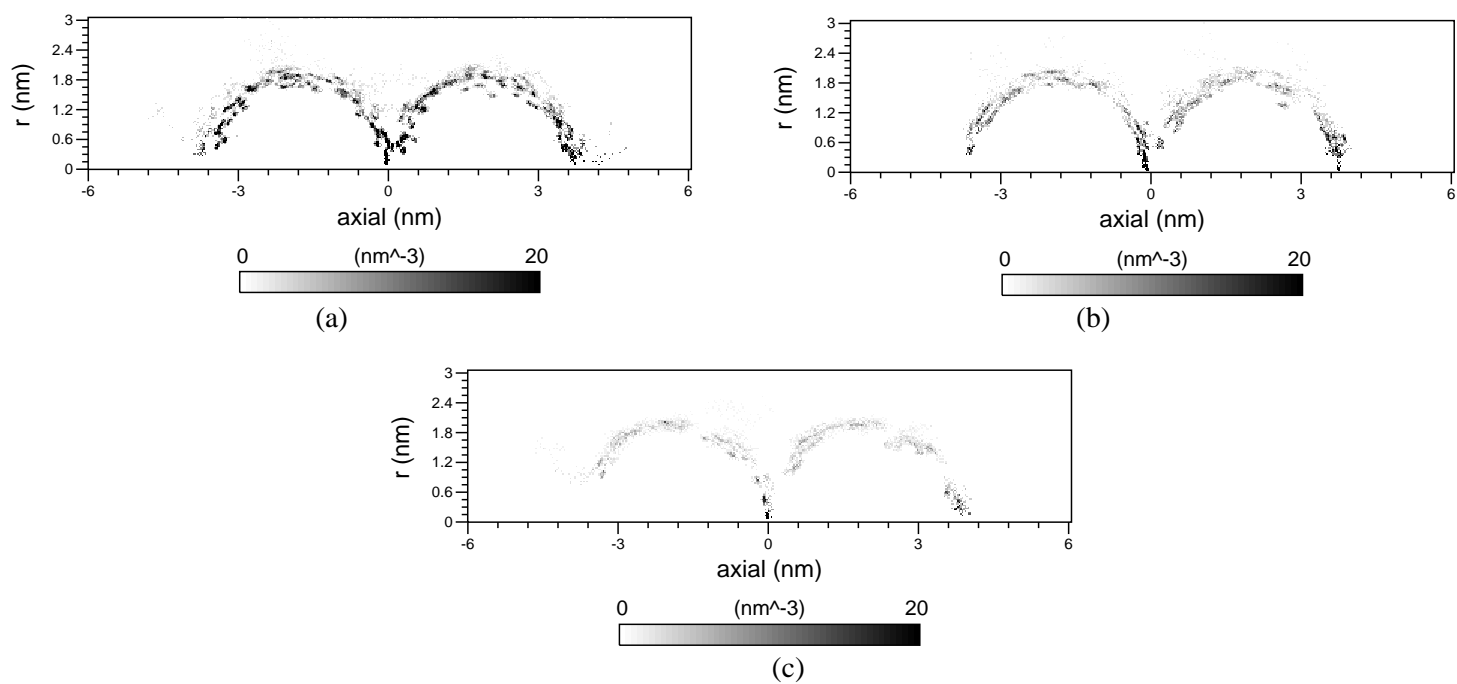

**Figure 2.** The axial-radial density plot of the distribution of sodium counter-ions around the pair of nanoparticles for the entire MD trajectory where the sub-figure (a) corresponds to the result for Si:Na$^+$ ratio of 5:1, (b) 10:1 and (c) 20:1. The axis corresponds to the line joining the centres of mass of the two particles. The same density scale is used for all three plots for ease of comparison.

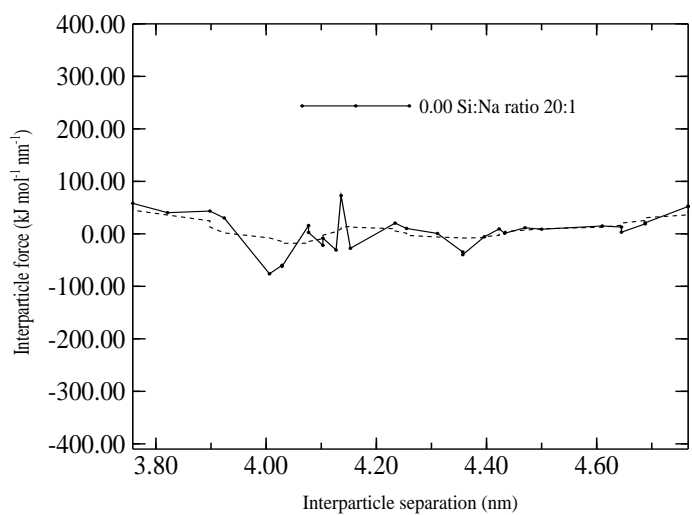
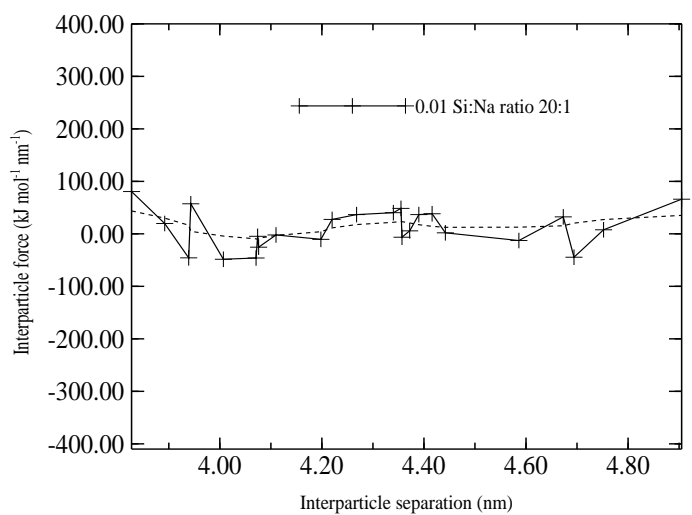
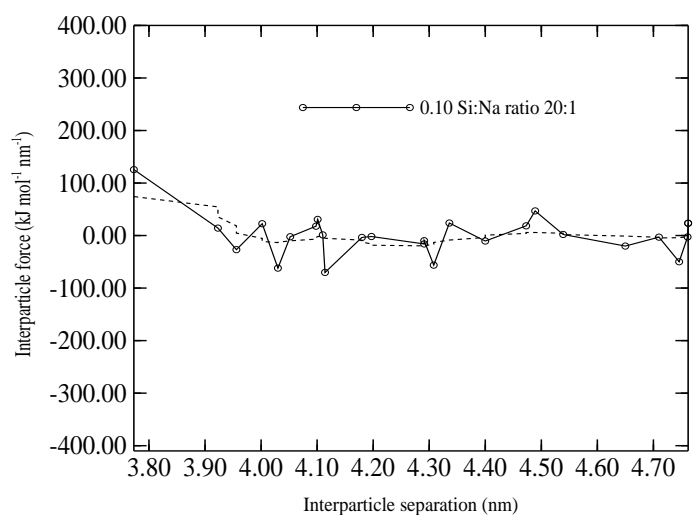
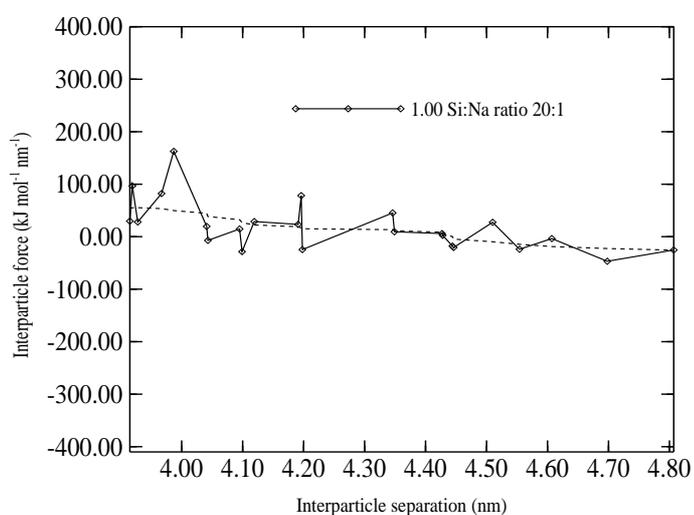

**Figure 3.** Plots of the inter-particle force vs. inter-particle separation for a pair of nanoparticles with diameter of 3.2 nm and Si:Na$^+$ ratio of 20:1 are shown in (a)-(d) for background sodium concentrations of 0.00, 0.01, 0.10 and 1.00 M respectively. Overlaid is the running average, shown as a dashed line.

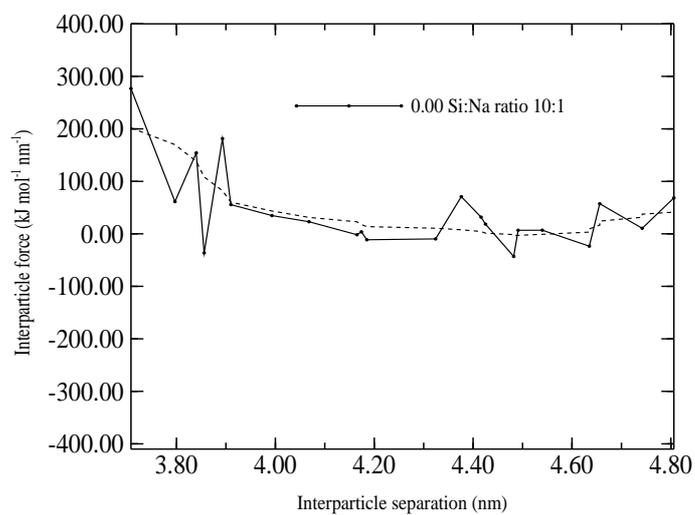
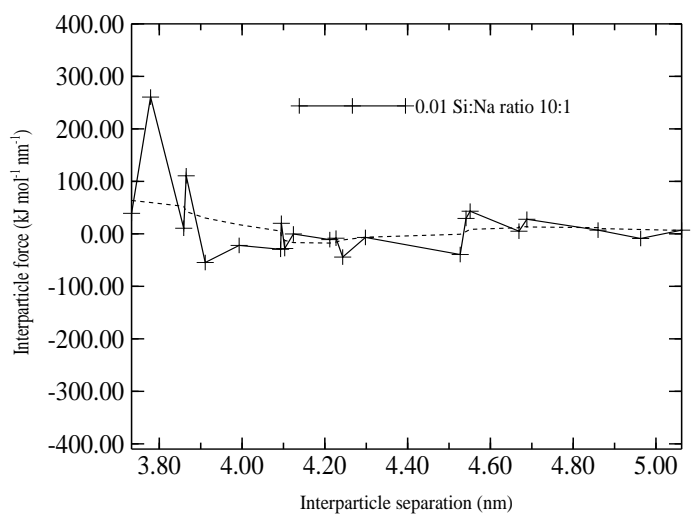
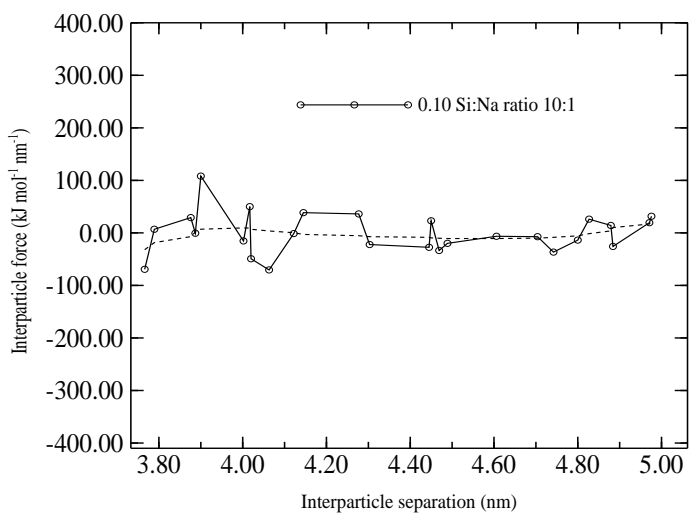
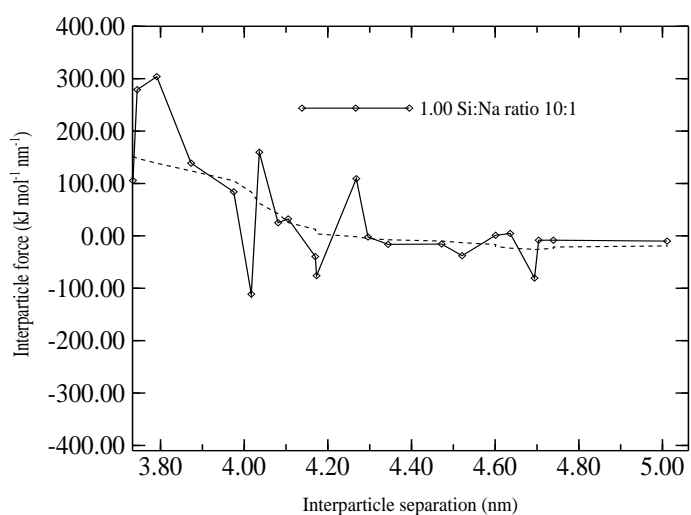

**Figure 4.** Plots of the inter-particle force vs. inter-particle separation for a pair of nanoparticles with diameter of 3.2 nm and Si:Na$^+$ ratio of 10:1 are shown in (a)-(d) for background sodium concentrations of 0.00, 0.01, 0.10 and 1.00 M respectively. Overlaid is the running average, shown as a dashed line.

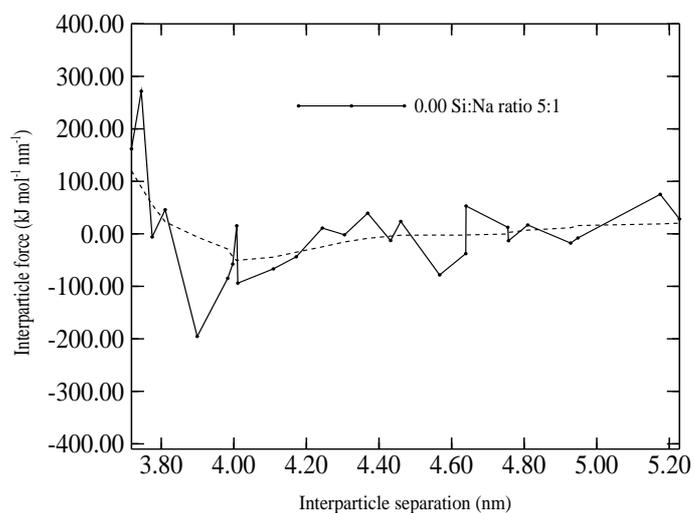
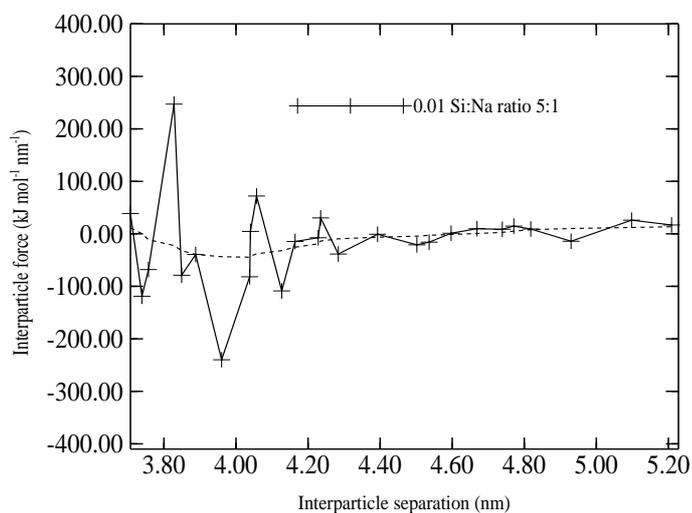
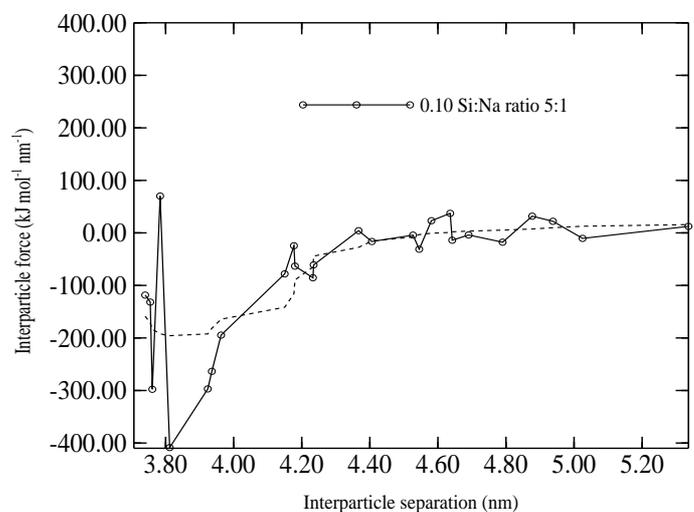
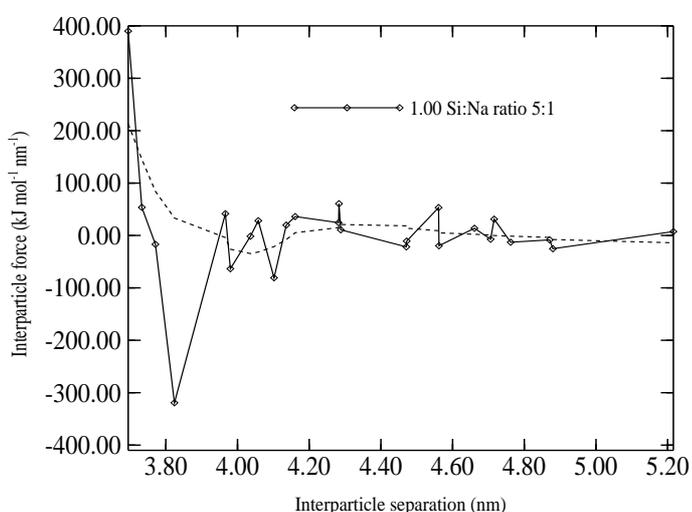

**Figure 5.** Plots of the inter-particle force vs. inter-particle separation for a pair of nanoparticles with diameter of 3.2 nm and Si:Na$^+$ ratio of 5:1 are shown in (a)-(d) for background sodium concentrations of 0.00, 0.01, 0.10 and 1.00 M respectively. Overlaid is the running average, shown as a dashed line.

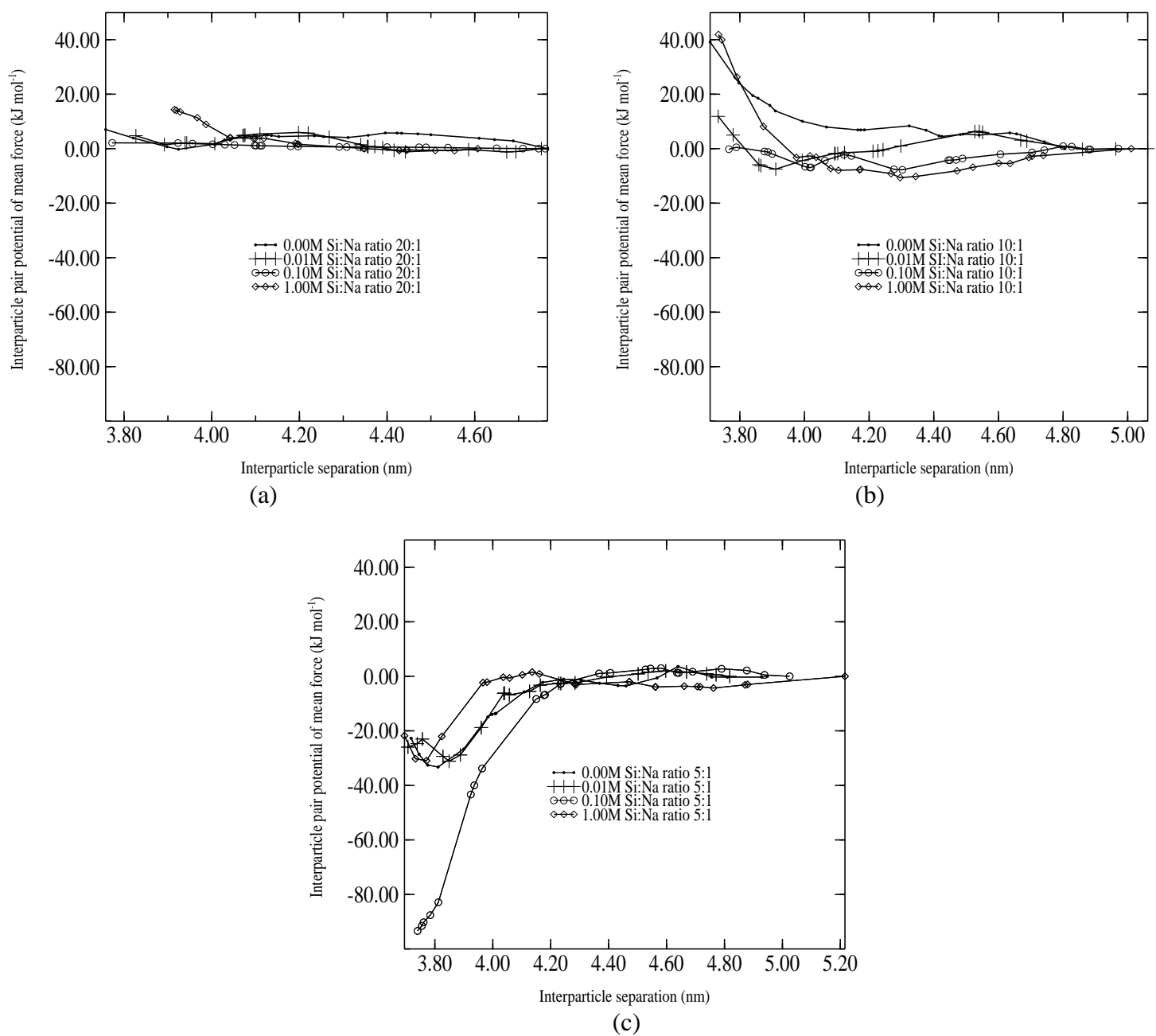

**Figure 6.** Plots of the inter-particle potential of mean force vs. inter-particle separation for a pair of nanoparticles with diameter of 3.2 nm and Si:Na$^+$ ratio of 20:1,10:1 and 5:1 are shown in (a),(b) and (c) respectively for the four background sodium concentrations of 0.00, 0.01, 0.10 and 1.00 M.